\documentclass[a4paper,12pt]{article} 
\pdfoutput=1 

 \usepackage{amsmath}
\usepackage{color}
\usepackage{amssymb}
\usepackage{bm}
\usepackage{graphicx}
\begin{document}
\titlepage

\today

  \begin{flushright}
   {\bf
  \begin{tabular}{l}
OCHA-PP-338\\
OCU-PHYS 439
  \end{tabular}
   }
  \end{flushright}

\vspace*{1.0cm}
\baselineskip 18pt
\begin{center}
{\Large \bf 
Perturbative unification of gauge couplings in supersymmetric $E_6$
 models
} 

\vspace*{1.0cm} 

{\bf 
Gi-Chol Cho$^a$, 
Nobuhito Maru$^b$, 
Kaho Yotsutani$^a$
}

\vspace*{0.5cm}
$^a${\em Department of Physics, Ochanomizu University, Tokyo 112-8610, Japan}\\
$^b${\em Department of Mathematics and Physics, Osaka City University, 
Osaka 558-8585, Japan}\\ 
\end{center}

\vspace*{1cm}

\baselineskip 18pt
\begin{abstract}
\noindent
 We study gauge coupling unification in supersymmetric $E_6$ models 
 where an additional $\mathrm{U}(1)'$ gauge symmetry is broken near the
 TeV scale and a number of exotic matter fields from the $\bm{27}$ 
 representations have $O(\mathrm{TeV})$ mass. 
  Solving the 2-loop renormalization group equations of gauge couplings
 and a kinetic mixing coupling between the $\mathrm{U}(1)'$ and 
 $\mathrm{U}(1)_Y$ gauge fields, we find that the gauge couplings fall
 into the non-perturbative regime below the GUT scale.
 We examine threshold corrections on the running of gauge couplings
 from both light and heavy ($\sim$ GUT scale) particles and 
 show constraints on the size of corrections to achieve 
 the perturbative unification of gauge couplings. 
\end{abstract}

\newpage
Grand unification of electromagnetic, weak and strong interactions is an
attractive feature of minimal supersymmetric standard model (MSSM). 
In grand unified theories (GUT), 
all matter fields in the MSSM are embedded in some larger representations
of a certain GUT group such as SU(5).
Among candidates of GUT groups, $E_6$ is known as a gauge group 
which is anomaly free, and each generation of quarks, leptons and Higgs 
superfields are embedded into one representation, i.e., $\bm{27}$ 
(for a review, see~\cite{Hewett:1988xc}).
The $E_6$ group could be decomposed as follows at the GUT scale: 
\begin{eqnarray}
 E_6 &\supset& {\rm SO(10)} \times {\rm U(1)}_\psi 
\supset 
{\rm SU(5)} \times {\rm U(1)}_\chi \times {\rm U(1)}_\psi. 
\label{e6breaking}
\end{eqnarray}
%%%
From phenomenological point of view, 
it is often expected that one of two extra U(1) symmetries (hereafter we 
call it $\mathrm{U}(1)'$) remains unbroken until the TeV scale. 
Then, many extra particles beyond the MSSM in $\bm{27}$ may have the mass of
order TeV. 
%para

%
In a certain extension of the MSSM, three gauge couplings are still 
unified at the unification scale $m_\mathrm{GUT}\simeq 2\times
10^{16}~\mathrm{GeV}$ in the leading order 
if the extra charged fields beyond the MSSM are
embedded into a vector-like pair of complete multiplet of SU(5), e.g., 
$\bm{5}+\overline{\bm{5}}$ or $\bm{10}+\overline{\bm{10}}$. 
The number of such extra fields, however, is constrained from the 
perturbativity of gauge couplings~\cite{Hempfling:1995rb}. 
The $\bm{27}$ representation in $E_6$ contains one generation of 
quark/lepton superfields, an extra pair of $\bm{5}+\overline{\bm{5}}$ 
and two SM singlet.
The Higgs superfields whose scalar components break the electroweak
symmetry should come from some other representations.
The supersymmetric (SUSY) $E_6$ models,
therefore, have at least three pairs of $\bm{5}+\overline{\bm{5}}$ in
addition to the MSSM fields as its low-energy spectrum.
%%%

%%%
In this paper, we study the gauge coupling unification in 
SUSY-$E_6$ models taking account of the kinetic mixing between
$\mathrm{U}(1)_Y$ and $\mathrm{U}(1)'$ beyond the leading order of
renormalization group equations (RGE)
\footnote{
The running of gauge couplings with the kinetic mixing in SUSY-$E_6$
models in the 1-loop level has been studied in
refs.~\cite{Dienes:1996du,Cho:1998nr,Rizzo:1998ut}.
}. 
We solve the 2-loop RGE of three ($\mathrm{SU}(3)_C$,
$\mathrm{SU}(2)_L$, $\mathrm{U}(1)_Y$) gauge couplings, 
the $\mathrm{U}(1)'$ gauge couplings and the 
$\mathrm{U}(1)_Y$-$\mathrm{U}(1)'$ kinetic mixing couplings.
Because of a number of extra particles beyond the MSSM, 
the running of gauge couplings in the SUSY-$E_6$ models are
asymptotic non-free, and the gauge couplings at the GUT scale is no 
longer perturbative. 
We examine constraints on the threshold corrections of light ($\sim 
$ TeV scale) and heavy ($\sim$~GUT scale) particles
from the perturbativity and experimental data. 
%para

%
We first briefly review the SUSY-$E_6$ models.
As is already mentioned, a linear combination of $\mathrm{U}(1)_\chi$ 
and $\mathrm{U}(1)_\psi$ in (\ref{e6breaking}) 
is assumed to remain in low-energy and the gauge boson $Z'$ is 
parametrized as 
\begin{eqnarray}
 Z' &=& Z_\chi \cos\beta + Z_\psi \sin\beta. 
\end{eqnarray}
As shown in Table~\ref{e6_beta}, 
there are some variants of $E_6$ models 
corresponding to the value of mixing angle $\beta$.
The couplings of matter fields and $Z'$ boson are trivial in both
$\chi$- and $\psi$-models\footnote{
In the $\psi$-model, 
$\sqrt{72/5}Q' =(1,-2,4)$ for $\bm{16},\bm{10},\bm{1}$
representations, respectively. 
In the $\chi$-model, 
$2\sqrt{6}Q'= (-1,3,-5), (2,-2)$ and 0 for
$(\bm{10},\overline{\bm{5}},\bm{1})$,
$(\bm{5},\overline{\bm{5}})$ and $\bm{1}$ representations. 
}, and those in the other models are given by a
linear combination of couplings in $\chi$- and $\psi$-models. 
In the following study, we adopt the $\eta$-model as a typical example, 
which is known as a consequence of direct breaking of $E_6$ into a
rank-5 group~\cite{Ellis:1986yg, Ellis:1985yc}.
Note that, only in the $\nu$-model, 
the right-handed neutrino $\nu^c$ could be gauge singlet under both the 
SM and $\mathrm{U}(1)'$ groups~\cite{Ibanez:1986si, Matsuoka:1986ie}. 
%---------------
\begin{table}[t]
 \begin{center}
  \begin{tabular}{|c|c|c|c|c|} \hline
   model &$\chi$ &$\psi$ &$\eta$ &$\nu$ \\ \hline
   $\beta$&0 &$\pi/2$ &$\tan^{-1}(-\sqrt{5/3})$ &$\tan^{-1}(\sqrt{15})$ \\ \hline 
  \end{tabular}
 \end{center}
\caption{various $E_6$ models versus the mixing angle $\beta$}
  \label{e6_beta}
\end{table}
%---------------
%para

%
The fundamental representation $\bm{27}$ in $E_6$ can be decomposed 
into representations in SO(10) and SU(5) as follows: 
\begin{eqnarray}
 \bm{27}
=\left\{\bm{16}+\bm{10}+\bm{1}\right\}_\mathrm{SO(10)}
=\left\{\left(\bm{10}+\bar{\bm{5}}+\bm{1}\right)
+\left(\bm{5}+\bar{\bm{5}}\right)
+ \bm{1}\right\}_\mathrm{SU(5)}. 
\end{eqnarray}
The $\mathrm{U}(1)'$ charge $Q'$ of all the matter fields in a $\bm{27}$
representation for the $\eta$-model is summarized in
Table~\ref{eta_charge}. The normalization of $\mathrm{U}(1)'$ charge
follows that of the hypercharge.
The $\mathrm{U}(1)'$ symmetry breaking could occur at near the weak
scale radiatively.
Discussions on this issue can be found, e.g.,
in refs.~\cite{Cvetic:1997ky,Langacker:1998tc}.
%---------------------
\begin{table}[t]
 \begin{center}
  \begin{tabular}{|c|ccc|cc|c|cc|cc|c|}  \hline
SO(10) & \multicolumn{6}{|c|}{$\bm{16}$} & \multicolumn{4}{|c|}{$\bm{10}$}&$\bm{1}$
\\ \hline
SU(5) &\multicolumn{3}{|c|}{$\bm{10}$}
&\multicolumn{2}{|c|}{$\bar{\bm{5}}$} &$\bm{1}$
						   &\multicolumn{2}{|c|}{$\bm{5}$}
							   &\multicolumn{2}{|c|}{$\bar{\bm{5}}$}&$\bm{1}$
											   \\ \hline
   &$Q$ &$u^c$ & $e^c$&$L$ &$d^c$ &$\nu^c$ &$H_u$ &$D$ &$H_d$ &$\bar{D}$
										   &$S$ \\ \hline 
$Y$ &$\frac{1}{6}$&$-\frac{2}{3}$ & $1$&$-\frac{1}{2}$ &$\frac{1}{3}$
					   &$0$ &$\frac{1}{2}$ &$-\frac{1}{3}$
								   &$-\frac{1}{2}$ &$\frac{1}{3}$ &0 \\
   \hline 
$Q'$ &\multicolumn{3}{|c|}{$-\frac{1}{3}$}
&\multicolumn{2}{|c|}{$\frac{1}{6}$} &$-\frac{5}{6}$
						   &\multicolumn{2}{|c|}{$\frac{2}{3}$}
							   &\multicolumn{2}{|c|}{$\frac{1}{6}$}&$-\frac{5}{6}$
											   \\ \hline
  \end{tabular}
 \end{center}
\caption{
The hypercharge $Y$ and the ${\rm U(1)'}$ charge $Q'$ of 
 all the matter fields in a {\bf 27} for the $\eta$ model. 
The value of $Q'$ follows the hypercharge normalization. }
\label{eta_charge}
\end{table}
%%%-------

%
The Lagrangian of neutral gauge bosons
($A^0$ which is a photon in the SM, $Z$ and $Z'$)  
is given by
\begin{eqnarray}
\mathcal{L}&=&
-\frac{1}{4}Z_{\mu\nu}Z^{\mu\nu}
-\frac{1}{4}Z'_{\mu\nu}Z'^{\mu\nu}
-\frac{\sin\chi}{2}B_{\mu\nu}Z'^{\mu\nu}
-\frac{1}{4}A^0_{\mu\nu}A^{0\mu\nu}
\nonumber \\
 &&+ m_{ZZ'}^2 Z_\mu Z'^\mu
  +\frac{1}{2}m_Z^2 Z_\mu Z^\mu
    +\frac{1}{2}m_{Z'}^2 Z'_\mu Z'^\mu, 
\end{eqnarray}
where $V_{\mu\nu}=\partial_\mu V_\nu -\partial_\nu V_\mu$ for a gauge
boson $V$. 
The parameter $\chi$ denotes the kinetic mixing angle between the
hypercharge gauge boson $B$ and the $\mathrm{U}(1)'$ gauge boson $Z'$.   
The mass eigenstates $(Z_1, Z_2, A)$ are obtained from the gauge
eigenstates $(Z, Z',A^0)$ via 
the mass and kinetic mixing angles $\xi$ and $\chi$, respectively (see, 
ref.~\cite{Cho:1998nr}).
In the limit of $\xi=0$, 
a shift of the $\mathrm{U(1)}'$ charge to the fermions is found as 
\begin{eqnarray}
 \mathcal{L}=\overline{\psi}\gamma^\mu
  \left\{	e Q A_\mu + g_Z (I_3 - Q \sin^2\theta_W) Z_{1\mu}
   + \frac{g_E}{\cos\chi}\left(
						  Q' - Y \frac{g_Z}{g_E}\sin\theta_W \sin\chi 
						 \right) Z_{2\mu}
				 \right\} \psi.
\label{offd}
\end{eqnarray}
The ``off-diagonal'' gauge coupling $g_{1E}$ and the kinetic mixing 
parameter $\delta$ are defined as 
\begin{eqnarray}
 g_{1E}&\equiv&-g_Z \sin\theta_W \sin\chi,
  \\
 \delta &\equiv& \frac{g_{1E}}{g_E}.
  \label{eq:delta}
\end{eqnarray}
Then, in the limit of $\xi=0$, the coupling of $Z_2$ boson to
the SM fermions are simply given by  
\begin{eqnarray}
 \tilde{Q}' &=& Q' + Y \delta. 
\end{eqnarray}
Note that, in this limit, the couplings of $Z_2$ to 
leptons in the $\eta$-model vanish when $\delta=1/3$.  
Next we discuss the gauge coupling unification in SUSY-$E_6$ models.
It should be noted that three gauge couplings are not unified in the
SUSY-$E_6$ models with three generations of $\bm{27}$ at the TeV scale.
This is because it does not much the unification condition on extension
of the MSSM by introducing couples of $\bm{5}+\overline{\bm{5}}$.  
For the gauge coupling unification, at least a pair of SU(2) doublet 
($\bm{2}+\bar{\bm{2}}$) should be added to the particle 
spectrum~\cite{Dienes:1996du}.
The origin of the additional pair of $\mathrm{SU}(2)$ doublet could be
either $\bm{27}+\overline{\bm{27}}$ or $\bm{78}$ in $E_6$.
It should be explained why $\bm{2}+\overline{\bm{2}}$ remains 
massless while the others decouple in a large representation. We will
return this point and mention some possibilities later. 
In the following study, we take the $\mathrm{U}(1)'$ charge of 
the additional $\mathrm{SU}(2)$ doublet to be $\pm\frac{1}{6}$, i.e.,
same with $L$ of $\bm{27}$ and its counter partner of 
$\overline{\bm{27}}$
\footnote{
The other choices are $H_d$ of $\bm{27}$ or $\overline{H_u}$ of
$\overline{\bm{27}}$.}.  
We also study the model of Babu et al.~\cite{Babu:1996vt} where two
pairs of $\bm{2}+\overline{\bm{2}}$ from $\bm{78}$ and a pair of 
$\bm{3}+\overline{\bm{3}}$ from $\bm{27}+\overline{\bm{27}}$ are added
to the $\eta$-model in order to achieve the 
quasi-leptophobity ($\delta 
\sim \frac{1}{3}$) through the 1-loop RGE. 
We refer this model as the $\eta_\mathrm{BKM}$-model in the rest of this
paper.
A similar study on gauge coupling unification in SUSY-$E_6$ models has
been presented in ref.~\cite{King:2007uj} focusing on the $\chi$-model. 
The authors in ref.~\cite{King:2007uj} neglected the kinetic mixing
parameter $\delta$ in their analysis  
because $\delta$ generated radiatively through the 1-loop RGE is quite
small. On the other hand, we investigate the $\eta$- and
$\eta_\mathrm{BKM}$-models taking account of the 2-loop RGE of kinetic 
mixing parameters as well as gauge couplings since, as mentioned above,
$\delta$ could become sizable in these models and a $Z'$ boson could be
leptophobic which is phenomenologically attractive. 
%%%----- 

%
The RGE of gauge couplings $g_i~(i=1\sim 5)$ is given by 
\begin{eqnarray}
 \frac{dg_i}{dt}&=&\beta_i^{(1)}+\beta_i^{(2)}, 
  \\
  t &\equiv&\ln \mu,
\end{eqnarray}
where $\mu$ stands for the renormalization scale, and 
$\beta_i^{(1)}$ and $\beta_i^{(2)}$ denotes the $\beta$-function
in 1- and 2-loop levels, respectively. 
We adopt the SU(5) normalization for the U(1) couplings and charges,
i.e.,
\begin{eqnarray}
 g_1 &=& \sqrt{\frac{5}{3}}g_Y,~~
  g_4 = \sqrt{\frac{5}{3}}g_E,~~
  g_5 = \sqrt{\frac{5}{3}}g_{1E},
  \\
 Q_1 &=& \sqrt{\frac{3}{5}}Y, ~~
 Q_E = \sqrt{\frac{3}{5}} Q'. 
\end{eqnarray}
The 1-loop part of the $\beta$ functions is given by\footnote{
The RGE given in this paper is based on the interactions in
eq.~(\ref{offd}).   
We note here that the RGE with the kinetic mixing between two
$\mathrm{U}(1)$ in a most general (``symmetric'') basis 
has been given in ref.~\cite{delAguila:1988jz}. }
\begin{eqnarray}
\beta^{(1)}_1 
 &=&
  \frac{1}{  16\pi^2}
		 b_1 g_1^3, 
\\
\beta^{(1)}_4
&=&
\frac{1}{  16\pi^2}
\left\{
g_4
\left(
b_E g_4^2 + b_1 g_5^2 
+ 2 b_{1E} g_4 g_5
	   \right)
\right\}, 
\\
\beta^{(1)}_5
&=&
\frac{1}{  16\pi^2}
\left(
b_E g_5 g_4^2
+ b_1 g_5^3
+ 2 b_1 g_1^2 g_5
+ 2 b_{1E} g_1^2 g_4
+ 2 b_{1E} g_5^2 g_4
\right), 
\\
  \beta^{(1)}_N
 &=&
\frac{1}{  16\pi^2} 
 b_N g_N^3 ~~~\mbox{(for $N$=2,~3)}. 
   \label{trr}
\end{eqnarray}
The coefficients $b_1, b_E$ and $b_{1E}$ are given by
\begin{eqnarray}
 b_1=\mathrm{Tr}(Q_1^2),~~
  b_E=\mathrm{Tr}(Q_E^2),~~
b_{1E}=\mathrm{Tr}(Q_1 Q_E), 
\end{eqnarray}
while the coefficient of r.h.s. in eq.~(\ref{trr}) is given by 
\begin{eqnarray}
 b_N &=& \sum  T(N) - 3 C_2(G), 
\end{eqnarray}
where
\begin{eqnarray}
 T(N)=\frac{1}{2},~~
  C_2 (N)= \frac{N^2 -1}{2N}, ~~
  C_2 (G)=N, 
\end{eqnarray}
for a fundamental representation $N$ and an adjoint representation $G$ 
of $\mathrm{SU}(N)$.
It should be summed over all charged fields under $\mathrm{SU}(N)$. 
The coefficients $b_i~(i=1,2,3), b_E$ and $b_{1E}$ in the $\eta$-
and $\eta_\mathrm{BKM}$-model are summarized in Table~\ref{tab1}, 
where the factor $a$ denotes the $\mathrm{U}(1)'$ charge of additional 
$\mathrm{SU}(2)$ doublets. 
The 1-loop RGE can be solved easily by assuming $g_1=g_2=g_3=g_4$ and
$\delta =0~(g_5=0)$ at the GUT scale. 
The magnitude of extra U(1) coupling $g_E$ and the kinetic mixing
parameter $\delta~(=g_{1E}/g_E)$ at the $m_Z$ scale are $g_E/g_Y=1.03$
and $\delta=0.018$ for the $\eta$-model with $a=1/6$ while
$g_E/g_Y=0.86$ and $\delta=0.29$ for the
$\eta_\mathrm{BKM}$-model, where the $\mathrm{U}(1)_Y$ gauge coupling
$g_Y$ is fixed at $g_Y=0.36$~\cite{Cho:1998nr}. 
We note that the kinetic mixing parameter $\delta$ in the 
  $\eta_\mathrm{BKM}$-model is close to the leptophobity condition,
  $\delta = 1/3$.  
The coefficients of 1-loop $\beta$-functions 
summarized in Table~\ref{tab1} tell us that gauge couplings $g_1 \sim 
g_4$ are asymptotically non-free and the running of gauge couplings is
expected to be affected by taking account of the 2-loop contributions.  
%para 

%
The 2-loop contributions to the RGE for $g_1, g_4$ and $g_5$ are
summarized as
\begin{eqnarray}
 %\frac{d}{dt}g_1
 (16\pi^2)^2
\beta^{(2)}_1
&=&
  4 Q_1^4 g_1^5
+ 4 Q_1^4 g_1^3 g_5^2 
+ 8 Q_1^3Q_E g_1^3 g_5 g_4
+ 4 Q_1^2Q_E^2 g_1^3 g_4^2
\nonumber \\
 &&
+ 
\sum_{N=2,3}
4C_2(N) Q_1^2 g_1^3 g_N^2,
\label{2loop1}
\end{eqnarray}
\begin{eqnarray}
(16\pi^2)^2
\beta^{(2)}_4
&=&
  4 Q_E^4 g_4^5 
+ 16 Q_1 Q_E^3 g_5 g_4^4 
+ 4 Q_1^2 Q_E^2 g_1^2 g_4^3 
+ 24 Q_1^2 Q_E^2 g_5^2 g_4^3 
\nonumber \\[2mm]
&&
+ 8 Q_1^3 Q_E g_1^2 g_5 g_4^2 
+ 16 Q_1^3 Q_E g_5^3 g_4^2 
+ 4 Q_1^4 g_1^2 g_5^2 g_4
+ 4 Q_1^4 g_5^4 g_4
\nonumber \\[2mm]
&&
+\sum_{N=2,3} 
C_2(N)\left(
  4 Q_E^2 g_4^3
+ 8 Q_1 Q_E g_5 g_4^2
+ 4 Q_1^2 g_5^2 g_4
\right)g_N^2,
\label{2loop4}
\end{eqnarray}

\begin{eqnarray}
(16\pi^2)^2
\beta^{(2)}_5
&=& 
  4 Q_1^4 g_5^5 
+ 16 Q_1^3 Q_E g_4 g_5^4 
+ 12 Q_1^4 g_1^2 g_5^3
+ 24 Q_1^2 Q_E^2 g_4^2 g_5^3 
\nonumber \\[2mm]
&&
+ 32 Q_1^3 Q_E g_1^2 g_4 g_5^2 
+ 16 Q_1 Q_E^3 g_4^3 g_5^2 
+ 8 Q_1^4 g_1^4 g_5
\nonumber \\[2mm]
&&
+ 28 Q_1^2 Q_E^2 g_1^2 g_4^2 g_5
+ 4 Q_E^4 g_4^4 g_5
+ 8Q_1^3 Q_E g_1^4 g_4
+ 8 Q_1 Q_E^3 g_1^2 g_4^3 
\nonumber \\[2mm]
&&
+\sum_{N=2,3} C_2(N) g_N^2
\left( 8 Q_1^2 g_1^2 g_5 
+ 4 Q_1^2 g_5^3 
+ 8 Q_1 Q_E g_1^2 g_4
\right.
\nonumber \\[2mm]
&&
\left.
+ 8 Q_1 Q_E g_5^2 g_4
+ 4 Q_E^2 g_5 g_4^2 
\right), 
\label{2loop5}
\end{eqnarray}
where the trace over all charged fields under the gauge groups are
understood. 
The explicit values of $(Q_1^4, Q_1^3 Q_E, Q_1^2 Q_E^2, Q_1 Q_E^3,
Q_E^4)$ in each model are summarized in Table~\ref{tab2}. 
\begin{table}[t]
\begin{center}
\begin{tabular}{cccccc}
 &$b_1$&$b_2$&$b_3$&$b_E$&$b_{1E}$\\ \hline \hline 
$\eta$
 &$\frac{48}{5}$&$4$&$0$&$9+\frac{12}{5}a^2 $& $-\frac{6}{5}a$
\\[2mm]
$\eta_\mathrm{BKM}$&$\frac{53}{5}$&$5$&$1$&$\frac{77}{5}$&$-\frac{16}{5}$ 
\end{tabular}
 \caption{Summary of coefficients $b_i$ in both $\eta$- and
 $\eta_\mathrm{BKM}$ models.
The $\mathrm{U}(1)'$ charge of additional pair of $\mathrm{SU}(2)$ is
 denoted by $a$. We take $a=\frac{1}{6}$ in our analysis. 
 }
 \label{tab1}
\end{center}
\end{table}
\begin{table}[h]
\begin{center}
\begin{tabular}{cccccc}
 &$Q_1^4$&$Q_1^3 Q_E$&$Q_1^2 Q_E^2$&$Q_1 Q_E^3$&$Q_E^4$\\ \hline \hline 
$\eta$& $\frac{117}{50}$& $-\frac{9}{50}a$&
			 $\frac{3}{4}+\frac{9}{25}a^2$&$-\frac{18}{25}a^3$&$\frac{9}{4}+\frac{36}{25}a^4 $ \\[2mm]
$\eta_\mathrm{BKM}$ &$\frac{737}{300}$
&$-\frac{31}{75}$&$\frac{473}{300}$&$-\frac{124}{75}$&$\frac{1667}{300}$
\\
\end{tabular}
\end{center}
 \caption{
 Explicit values of factors in eqs.~(\ref{2loop1}), 
 (\ref{2loop4}) and (\ref{2loop5}). 
We take $a=\frac{1}{6}$ in our study. 
 }
\label{tab2}
\end{table}

The 2-loop contributions to the 
RGE of non-abelian couplings $g_N~(N=2,3)$ is given as follows 
\begin{eqnarray}
 (16 \pi^2)^2 \beta^{(2)}_N &=&
  g_N^5\left[
		\left\{2 C_2 (G) + 4 C_2 (N)\right\} T(N) d(N')
		-6 \left\{C_2 (G)\right\}^2
			  \right]
  \nonumber \\
 &&+ g_N^3 g_{N'}^2
  \left\{    4 C_2 (N') T(N) d(N')  \right\}
  \nonumber \\
 &&+ g_N^3 \left(
			2 Q_1^2 g_1^2 + 2 Q_1^2 g_5^2 + 4Q_1 Q_E g_5 g_4 +
			2Q_E^2 g_4^2
		   \right) d(N'), 
\end{eqnarray}
where $g_{N'}$ and $d(N')$ denote the gauge coupling and the dimension
of charged fields of another non-abelian gauge group $\mathrm{SU}(N')$,
respectively. For example, $d(N')=3~(1)$ for a triplet (singlet) in
SU(3) for $N=2$. 
In RGE of $g_i$ ($i=1$-5), 
contributions from the Yukawa couplings of fermions at the 2-loop level
are not included. 
Since the Yukawa couplings of exotic fermions and the flavor mixings are
model dependent, including their effects increases the model parameters
and makes the analysis complicated. 
The impacts of the Yukawa couplings to our results will be discussed
later. 
%para

%
%%%---------
\begin{figure}[t]
\begin{center}
 \includegraphics[width=15cm,clip]{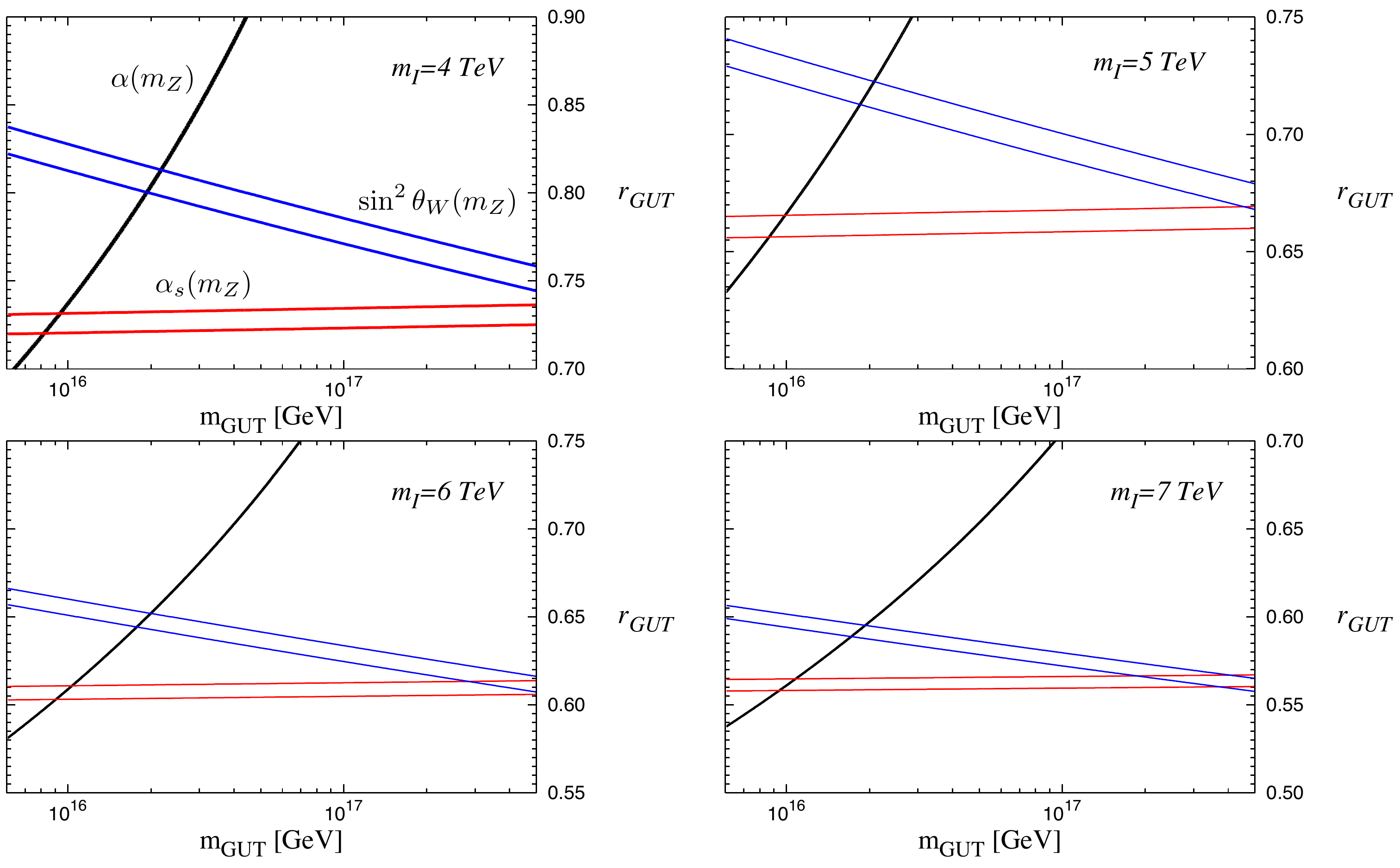}
 \caption{
 Predictions on $\alpha(m_Z)$ (black), $\alpha_s(m_Z)$ (red) and
 $\sin^2\theta_W (m_Z)$ (blue) in the $\eta$-model with
 $a=\frac{1}{6}$. 
 Each band stands for the region on
$(m_\mathrm{GUT}, r_\mathrm{GUT})$ plane which 
 satisfies $\Delta \chi^2 < 4$. 
 }
 \label{fig:eta}
\end{center}
\end{figure}
%%%---------
Next we test the gauge coupling unification in SUSY-$E_6$ models
numerically.
We solve the 2-loop RGE from the GUT scale taking the unification scale
$m_\mathrm{GUT}$ and the unified gauge coupling $\alpha_\mathrm{GUT}$ as
inputs. 
%In the analysis, we introduce a ratio of $\alpha_\mathrm{GUT}$ in 1- and
%2-loop levels as an input parameter: 
%\begin{eqnarray}
% r_\mathrm{GUT}\equiv \frac{\alpha_\mathrm{GUT}^\mathrm{2-loop}}
%  {\alpha_\mathrm{GUT}^{\mathrm{1-loop}}}. 
%\end{eqnarray}
%
In the analysis, we introduce a ratio of $\alpha_\mathrm{GUT}$ in 1- and
2-loop levels as; 
\begin{eqnarray}
 r_\mathrm{GUT}\equiv \frac{\alpha_\mathrm{GUT}^\mathrm{2-loop}}
  {\alpha_\mathrm{GUT}^{\mathrm{1-loop}}}, 
\end{eqnarray}
where $\alpha_\mathrm{GUT}^{\mathrm{1-loop}}$ is found by solving the
1-loop RGE, i.e., $1/\alpha_\mathrm{GUT}^{\mathrm{1-loop}}=3.17$. 
Then we obtain the input $\alpha_\mathrm{GUT} \equiv
\alpha_\mathrm{GUT}^\mathrm{2-loop}$ by varying $r_\mathrm{GUT}$.  
In general, the unification scale $m_\mathrm{GUT}$ is understood as a
scale where the $\mathrm{SU}(5)$ symmetry is broken to the SM gauge
group, and the $E_6$ 
symmetry breaking scale may be higher than $m_\mathrm{GUT}$. In our
study, however, we assume that the $E_6$ symmetry is broken at 
$m_\mathrm{GUT}$ since the $\eta$-model is obtained when $E_6$ is
directly broken to a rank 5 group as is already mentioned above. 
We also introduce the intermediate scale $m_I$ in which the threshold 
corrections by exotic particles beyond the MSSM in SUSY-$E_6$ models are
switched on, i.e., the $\beta$-functions in SUSY-$E_6$ model change to
MSSM at $m_I$. The mass scale of all MSSM particle are assumed to be
1~TeV.
Solving the 2-loop RGE with these input parameters, 
we compare the gauge couplings at the $m_Z$ scale with the experimental
values~\cite{Beringer:1900zz}
\begin{eqnarray}
 1/\alpha(m_Z)&=&127.944 \pm 0.014,
  \label{alpha}
  \\
 \alpha_s(m_Z)&=&0.1185\pm 0.0006,
  \label{alphas}
\\
 \sin^2\theta_W(m_Z)&=&0.23116 \pm 0.00012.
    \label{sn2w}
\end{eqnarray}
%
%para

\begin{figure}[t]
\begin{center}
         \includegraphics[width=15cm,clip]{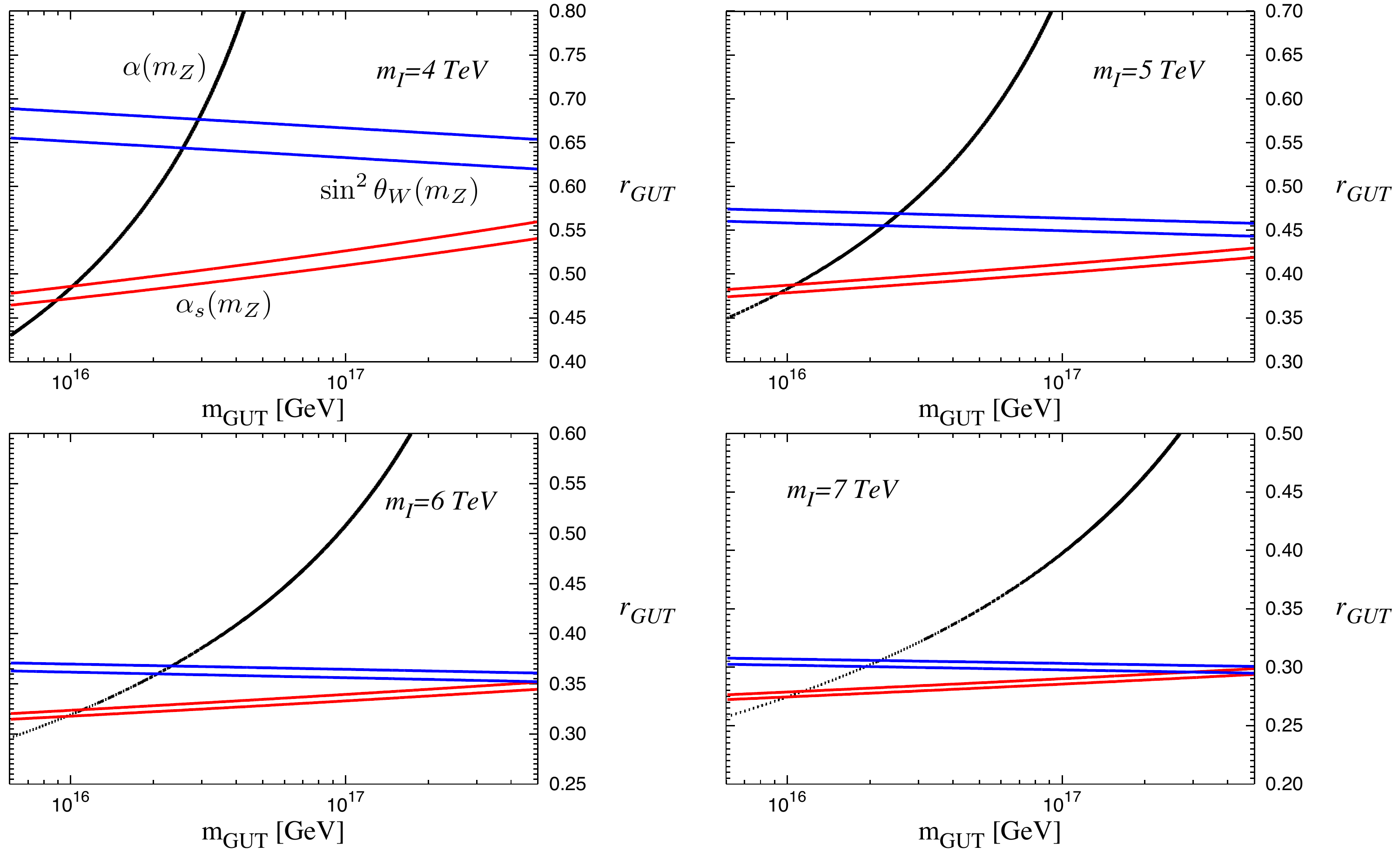}
 \caption{
 Same with those in Fig.~\ref{fig:eta} 
 but for the $\eta_{\rm BKM}$-model.
 }
  \label{fig:BKM}
\end{center}
\end{figure}

%
We give our results for $m_I=4,5,6$ and $7~\mathrm{TeV}$ in 
Figs.~\ref{fig:eta} ($\eta$-model with $a=1/6$) and \ref{fig:BKM} 
($\eta_\mathrm{BKM}$-model). 
These figures 
show the regions which satisfy $\Delta \chi^2<4$
for $\alpha(m_Z),~\alpha_s(m_Z)$ and $\sin^2\theta_W(m_Z)$ 
on the ($m_\mathrm{GUT},~r_\mathrm{GUT}$) plane 
by blue, red and black bands, respectively.
Three gauge couplings $(g_1, g_2, g_3)$ are successfully unified when 
three bands cross each other on the ($m_\mathrm{GUT}$, $r_\mathrm{GUT}$)
plane.
No such a crossing of three bands, however, is found in the figures. 
Discrepancies of three bands decreases as $m_I$ increases, since 
the running of gauge couplings coincides with that in the MSSM in the
limit of $m_I \to m_\mathrm{GUT}$.
The intermediate scale $m_I$ is, therefore, required to be high for 
the coupling unification of SUSY-$E_6$ models.
We note that the kinetic mixing parameter $\delta$ in eq.~(\ref{eq:delta}) is
highly suppressed as $\mathcal{O}(10^{-2}-10^{-3})$ in a wide range of
parameter space. 
%%%---------
%%%---------

%
Finally we discuss threshold corrections from the heavy particles whose
mass scale is around the GUT scale. We have mentioned particle spectrum
which is charged under the $\mathrm{U}(1)'$ gauge symmetry
i.e., 
three $\bm{27}$ representations and extra matters for the coupling 
unification.
We have so far not discussed, however, the heavy particles such as the
Higgs fields to break the $E_6$ symmetry.
Although such fields decouple from the light spectrum by getting the GUT
scale mass, these fields may contribute to the running of gauge
couplings near the GUT scale. 
Such corrections are called the heavy particle threshold corrections 
which are proportional to  $c\times \ln M/m_\mathrm{GUT}$ where $M$ is a
mass of the field and the coefficient $c$ is determined by the charge of
the field under $E_6$. 
Since we have not considered concrete heavy spectrum of SUSY-$E_6$ 
models which will decouple after the $E_6$ breaking, we estimate 
the magnitude of threshold corrections from 
the heavy particles required for
the successful coupling unification. 
%para

%
 Introducing a parameter $\Delta_i$ which accounts for the threshold
 corrections from the heavy particles, the gauge couplings at the $m_Z$ 
 scale can be expressed as 
\begin{eqnarray}
 \frac{1}{\alpha_i(m_z)}&=&
 \frac{1}{\alpha_i(m_z)_\mathrm{1+2 loops}} + \Delta_i. 
\label{eq:hth}
\end{eqnarray}
We perform the $\chi^2$-fit of $\Delta_i$ to the data of $\alpha(m_Z), 
\alpha_s(m_Z)$ and $\sin^2\theta_W(m_Z)$ in
eqs.~(\ref{alpha})-(\ref{sn2w})  
for the intermediate scale 
$m_I=5$ and $10~\mathrm{TeV}$, and results are summarized in 
Table~\ref{tb:heavy}.
As expected, $\Delta_i$ is required to be sizable for the coupling
unification when the intermediate scale $m_I$ is smaller. 
%%%--- 
\begin{table}[t]
 \begin{center}
	\begin{tabular}{r|rr}
   \multicolumn{2}{c}{$~~~\eta$-model}& \multicolumn{1}{c}{$\eta_\mathrm{BKM}$-model}\\ \hline
	  \multicolumn{3}{c}{$m_I=5~\mathrm{TeV}$} \\ \hline 
   $\Delta_1$   & $-1.013\pm 0.011$& $-0.990\pm 0.011$\\
   $\Delta_2$   & $-1.006\pm 0.016$& $-0.875\pm 0.016$\\
	$\Delta_3$   & $0.011\pm 0.043$& $0.494\pm 0.043$\\ \hline
	  \multicolumn{3}{c}{$m_I=10~\mathrm{TeV}$} \\ \hline 	
   $\Delta_1$   & $-0.677\pm 0.011$& $-0.542\pm 0.011$\\
   $\Delta_2$   & $-0.672\pm 0.016$& $-0.429\pm 0.016$\\
	$\Delta_3$   & $0.344\pm 0.043$& $0.948\pm 0.043$\\ \hline
	\end{tabular}
 \end{center}
 \caption{Constraints on the heavy particle threshold corrections
 $\Delta_i$ for $m_I=5$ and $10~\mathrm{TeV}$.
The correlation $\rho$ between the error in $\Delta_1$ and that in
 $\Delta_2$ is given as $\rho=-0.68$ in each case.
 }
 \label{tb:heavy}
\end{table}

To summarize, we have studied the gauge coupling unification in the 
SUSY-$E_6$ model taking account of both the
$\mathrm{U}_Y$-$\mathrm{U}(1)'$ mixing and the 2-loop contributions to
the RGE. The minimal model which maintains the coupling
unification consists of three generation of $\bm{27}$ and a pair of
$\bm{2}+\bar{\bm{2}}$. As an example, we focused on the $\eta$-model
which breaks the $E_6$ symmetry directly into the SM gauge group.
We also studied the $\eta_\mathrm{BKM}$-model where two 
$\bm{2}+\bar{\bm{2}}$ and one $\bm{3}+\bar{\bm{3}}$ are added to three
generations of $\bm{27}$. 
We found that results in the 1-loop RGE are significantly affected by 
2-loop corrections, and constraints on experimental measurements of 
$\alpha, \alpha_s$ and $\sin^2\theta_W$ at the $m_Z$ scale require the 
intermediate scale $m_I$ to be much higher than $O(1~\mathrm{TeV})$. 
We also obtained constraints on the size of threshold corrections from
heavy particles $\Delta_i$ to achieve the coupling unification.
A few comments are in order. 
Throughout our analysis, we neglected contributions from the Yukawa
couplings of fermions for simplicity. Since, in general, the Yukawa
couplings negatively contribute to the running of gauge couplings, they 
might affect the results if they are not negligible. In our analysis, we
have so far expressed the threshold corrections from heavy particles by
model independent parameters $\Delta_i$ in eq.~(\ref{eq:hth}).
The parameters $\Delta_i$, however, can be understood to represent 
a sum of contributions from the Yukawa couplings and the heavy threhold 
corrections, and the combinations of two contributions are constrained
by experimental data as shown in Table~\ref{tb:heavy}. 
We also comment on how additional massless $\bm{2}+\bar{\bm{2}}$ at the GUT
scale originated from $E_6$ multiplets. 
One way is the sliding singlet mechanism~\cite{Witten} or the missing
partner mechanism~\cite{missing}, which have been known as solutions 
to the doublet-triplet splitting problem in the $\mathrm{SU}(5)$ GUT. 
Another way is the mechanism in extra dimensional models~\cite{Kawamura}.  
Let us suppose a five dimensional model compactified on an orbifold $S^1/Z_2$ 
and the compactification scale is the GUT scale. 
We consider the case where the SM fields except for Higgs doublets are
localized on the fixed points while the gravity and the $E_6$ multiplets 
including Higgs fields propagate in the bulk.  
The $E_6$ gauge symmetry is broken by the boundary conditions ($Z_2$
parity).  
If only the Higgs doublet components $\bm{2}+\bar{\bm{2}}$ are assigned
to be $Z_2$ even and the others are $Z_2$ odd, then only 
$\bm{2}+\bar{\bm{2}}$ remains to be massless at the GUT scale and 
the others have at least the GUT scale masses.

%\section*{Acknowledgements}

%
\end{document}